\documentstyle[epsfig,12pt,preprint,tighten,aps]{revtex}
\begin{document}


\rightline{July 2001}
\rightline{To appear in Phys.Lett.B}
\vskip 2cm
\centerline{\large \bf  
A mirror world explanation for the Pioneer spacecraft anomalies?
}
\vskip 1.1cm
\centerline{R. Foot and R. R. Volkas,\footnote{E-mail address:
foot@physics.unimelb.edu.au, r.volkas@physics.unimelb.edu.au}}
\vskip .7cm
\centerline{{\it School of Physics}}
\centerline{{\it Research Centre for High Energy Physics}}
\centerline{{\it The University of Melbourne}}
\centerline{{\it Victoria 3010 Australia}}
\vskip 2cm

\centerline{Abstract}
\vskip 1cm
\noindent
We show that the anomalous acceleration
of the Pioneer 10/11 spacecraft can be explained
if there is some mirror gas or mirror dust 
in our solar system.

\vspace{2cm}
\noindent
\newpage

The unaccounted-for component of acceleration observed
for the Pioneer 10 and 11 spacecraft
presents an interesting scientific mystery \cite{det1,det}. 
These spacecraft, which are identical in design,
were launched in the early 1970's
with Pioneer 10 (11) approaching Jupiter (Saturn).
After these planetary rendezvous, 
the two spacecraft followed
hyperbolic orbits near the plane of the ecliptic
to opposite ends of the solar system with roughly
the same speed, which is now about 12 km/s.
The radiation pressure decreases quickly with
distance from the sun, and for distances greater
than 20 AU it is below $5 \times
10^{-8} \ cm/s^2$ allowing for a sensitive test for
anomalous forces in the solar system\cite{det1}. The
Pioneer 11 radio system failed in 1990 when it was about 30
AU away from the Sun, while Pioneer 10 is in better shape
and is about 70 AU away from the Sun (and still transmitting!).

Interestingly, careful and detailed studies of the motion
of Pioneer 10 and 11 have revealed
that the acceleration of {\it both} spacecraft 
is (or was in the case of Pioneer 11) anomalous and directed 
roughly towards the Sun \cite{det1,det}, with magnitude
\begin{equation}
a_p = (8.7 \pm 1.3)\times 10^{-8}\ \ cm/s^2\ .
\end{equation}
Many explanations have been
proposed, but all have been found wanting so far (for 
a review see \cite{det}). In this paper we 
point out that a cloud of mirror matter gas or dust in our solar
system could account for the observations by inducing a drag force.

We first briefly review the mirror matter idea.
One of the most natural candidates for a symmetry of nature is 
parity (i.e.\ left-right) symmetry, an improper Lorentz transformation.
While it is an established 
experimental fact that parity symmetry appears broken because of
the chiral asymmetry of weak interactions,
this actually does not 
exclude the possible existence of exact or unbroken parity symmetry in
nature. 
This is because parity (and also time reversal) can be exactly conserved 
if a set of mirror particles exist \cite{ly,flv}.  The idea is 
that for each ordinary particle, such as the photon, electron, proton
and neutron, there is a corresponding mirror particle, of exactly 
the same mass as the ordinary particle. The electromagnetic, weak
and strong forces are also doubled. 
In the modern language of gauge theories, the mirror particles 
are all singlets under the standard $G \equiv SU(3)\otimes SU(2)_L 
\otimes U(1)_Y$ gauge interactions. Instead the mirror
fermions interact with a set of mirror gauge particles:
the gauge symmetry of the theory is doubled
to $G \otimes G$. (The ordinary particles are, of 
course, singlets under the mirror gauge symmetry \cite{flv}.)
Parity is conserved because the mirror fermions experience
$V+A$ mirror weak interactions while the ordinary fermions 
experience the usual $V-A$ weak interactions. 
The two sectors are almost decoupled, 
but they interact
gravitationally and in general through other subtle effects such as 
the photon - mirror photon mixing we will exploit below.

Mirror protons and electrons are stable for the same reasons that 
ordinary protons and electrons are stable.
Mirror matter thus provides a candidate for the inferred dark
matter in the universe \cite{blin}. Several astrophysical
puzzles can potentially be resolved by mirror matter.
For example, observations of 
gravitational microlensing \cite{macho} in the galactic halo suggest
the presence of compact objects averaging roughly half a solar mass.
The most reasonable conventional candidates, white dwarfs,
pose serious phenomenological problems through the heavy element
background that should have been produced by the progenitor stars \cite{freese}.
Mirror compact objects are not subject to this objection \cite{ii}. Large
close-in extra-solar planets \cite{qel} came as a surprise, because they
could not have formed at their detected locations if they are made of
ordinary matter. It is possible they are instead made of mirror matter
\cite{mp} (the conventional alternative is that they formed further from
their stars and then migrated in).
The qualitative mirror image of such a system is an ordinary planet
orbiting a mirror star. Such hybrids could also exist. Indeed, we
have speculated that
the objects termed ``isolated
planetary mass objects'' \cite{zapatero}
may not be isolated at all, but rather they could
be ordinary planets orbiting invisible stellar companions composed of
mirror matter \cite{iso} (this can be tested by Doppler observations). The
last two examples illustrate the
likely segregation of ordinary and mirror matter. While hybrid systems
should exist, for phenomenological (and theoretical) reasons one
expects very uneven mixtures: mainly ordinary matter with a small
amount of mirror matter, and vice-versa. Our solar system could contain a
small amount of mirror matter (for bounds on mirror matter in the Earth see
Ref.\cite{iv1}).

While gravity is the predominant common interaction,
small non-gravitational interactions
are also possible and could be very important.  Due to
constraints from gauge symmetry, renormalizability and parity
symmetry it turns out that there are only three non-gravitational
ways in which
ordinary and mirror matter can interact with each other \cite{flv,flv2}. 
These are via photon - mirror photon
kinetic mixing, Higgs - mirror Higgs interactions and
via ordinary neutrino - mirror neutrino mass mixing
(if neutrinos have mass). While Higgs - mirror Higgs
interactions will be tested if or when the Higgs particle
is discovered \cite{flv2,iv2}, there is currently interesting evidence
for photon - mirror photon
kinetic mixing from the orthopositronium lifetime puzzle \cite{fg} and
also ordinary neutrino - mirror neutrino mass mixing
from the observed neutrino anomalies \cite{flv2,P}.

Since neutrino physics is in a state of flux, a short digression is
warranted:
If neutrinos have mass then a necessary consequence of
the parity symmetry of the theory is that each of the 
ordinary neutrinos $\nu_{e,\mu,\tau}$
oscillates maximally with its mirror 
partner $\nu'_{e,\mu,\tau}$. This provides a simple and predictive solution
to the solar and atmospheric neutrino anomalies
which is also compatible with the LSND \cite{lsnd} signal \cite{P}.
This mirror world solution predicted \cite{P,pred} the
energy independent recoil electron spectrum
observed for solar neutrinos at Super-Kamiokande \cite{sk}
as well as the observed $\sim 50\%$ flux reduction obtained
in the gallium experiments \cite{gal}. It also predicted
the maximal mixing observed in atmospheric neutrino
experiments \cite{P}.

It is true though that
the explanation of the neutrino anomalies does not
provide a perfect fit to all of the neutrino data.
In particular, the low Homestake result 
is about 3 sigma less than the predicted $\sim 50\%$
flux reduction.
(Homestake measures about one-third of the expected value for 
solar neutrinos while six other experiments measure about one-half). 
Recent SNO data disfavours 
$\nu_e \to \nu'_e$ oscillations at about the $3 \sigma$ level\cite{sno}.
This result arises from a comparison of SNO charged current
results with Super-Kamiokande electron elastic scattering measurements,
both of which are dominated by systematic uncertainties.
Finally some atmospheric data disfavour maximal $\nu_\mu \to \nu'_\mu$
oscillations
at about the $1.5 - 3 \sigma$ level depending on how the data is 
analysed \cite{skxx,foot2000}. A convincing test of the mirror
world explanation of the neutrino anomalies will be
provided soon by SNO's neutral/charged-current measurement. This should 
provide a solid result ($ > 7 \sigma$) one way or the other.

In field theory
photon - mirror photon kinetic mixing is described by the interaction 
\begin{equation}
{\cal L} = {\epsilon \over 2}F^{\mu \nu} F'_{\mu \nu},
\label{ek}
\end{equation}
where $F^{\mu \nu}$ ($F'_{\mu \nu}$) is the field strength 
tensor for electromagnetism (mirror electromagnetism).
This type of Lagrangian term is gauge invariant 
and renormalizable and can exist at tree level \cite{fh,flv}
or may be induced radiatively in models without $U(1)$ 
gauge symmetries (such as grand unified theories) \cite{bob,gl,cf}.
One effect of ordinary photon - mirror photon kinetic mixing
is to give the mirror charged particles a small electric
charge \cite{bob,gl,flv}. That is, they couple to ordinary photons with
electric charge $\epsilon e$.
It turns out that orthopositronium is peculiarly sensitive
to photon - mirror photon kinetic mixing\cite{gl} and
the anomalous vacuum cavity experiments suggest that
$\epsilon \simeq 10^{-6}$ at $5 \sigma$ 
level \cite{fg}.\footnote{This is large enough to be cosmologically
significant, because photon - mirror photon mixing would then bring the 
mirror sector into thermal equilibrium with the ordinary plasma prior
to the big bang nucleosynthesis (BBN) epoch \cite{cg}. This is consistent with
the recent microwave background anisotropy measurements \cite{cmbr,steen}, but it would 
suggest a modification of standard BBN such as a 
suitably large relic neutrino asymmetry \cite{nuasym}.}

Any mirror matter in our solar system may have formed
planets or small asteroid-sized objects and there may also
be some mirror gas or mirror dust\footnote{
Some possible effects of mirror planets in our solar system
have been discussed recently in Ref.\cite{sil}.} .
Collisions of mirror matter space bodies with
the Earth will result in observable effects
if the photon-mirror photon kinetic mixing is large enough.
The $\epsilon \simeq 10^{-6}$ figure suggested by the orthopositronium
anomaly is sufficiently large.
It has been argued in Refs.\cite{tung,puz} that various
observed events such as the Tunguska explosion
may have been due to the collision of the Earth 
with a mirror matter space body.

Now, the Pioneer 10/11 spacecraft are very sensitive probes
of mirror gas and dust in our solar system if $\epsilon \simeq 10^{-6}$.
Collisions of the spacecraft with mirror particles
will lead to a drag force which will slow down the
spacecraft\footnote{
It was originally thought that a drag force could not explain
the anomalies because of Ulysses data\cite{det1}, however
latter it was found that large systematic errors apparently
due to gas leaks made Ulysses data unreliable for a test
of the anomalous acceleration\cite{det3}.}. 
This situation of an ordinary matter body
(the spacecraft)
propagating though a gas of mirror particles is
dynamically the `mirror image' of a mirror matter space body
propagating through the atmosphere which was
considered in Ref.\cite{tung}.
For $\epsilon \approx 10^{-6}$ it turns out that the
relative momentum between the spacecraft and mirror
atoms is lost (up to random thermal motion) after the
mirror atoms penetrate a distance within the spacecraft of
roughly \cite{tung}
\begin{equation}
z 
\sim 0.1 \left({10^{-6} \over \epsilon}\right)^2 \left({v \over 10 
\ km/s}\right)^4 \  \ {\rm mm}. 
\label{DDD}
\end{equation}
Because the mirror atoms lose their relative momentum within the
spacecraft the drag force is of the usual form,
\begin{equation}
F_{drag} =  \rho_{mirror} A v^2 \ ,
\label{drag}
\end{equation}
where $\rho_{mirror}$ is the density of mirror particles
in the solar system in the path of the pioneers
and $v, A$ is the spacecraft speed and cross sectional
area respectively. 

The sign of the anomaly is consistent with a drag
force because the spacecraft are travelling in a direction
which is close to being radially outward from the Sun.
A drag force due to the interactions of 
ordinary matter cannot
explain the anomaly because of stringent constraints on its
density \cite{det}.
However the constraints on mirror matter in our solar system are
much weaker because of its invisibility as far as its interactions
with ordinary light is concerned.
Given the spacecrafts' mass of $241\ kg$ and cross sectional
area of about $5 \ m^2$\cite{det}, 
Eq.(\ref{drag}) can be re-written in the form:
\begin{equation}
a_{drag} \simeq 10^{-7}\left({\rho_{mirror} \over 4 \times 10^{-19} 
\ g/cm^3}\right)
\left( {A \over 5 m^2}\right)\left({v \over 12 km/s}\right)^2
cm/s^2 \ .
\end{equation}
Thus, the anomalous acceleration measurements of the Pioneer
10/11 spacecraft suggest a density of mirror matter
in the solar system of about $\approx 4 \times 10^{-19}\ g/cm^3$.

The approximately constant nature of the Pioneer anomalies 
requires that the
density be roughly constant in the plane of the ecliptic
to within about $20-30\%$
between $25-60$ AU. We cannot prove that mirror matter
will form such a configuration. It depends on the abundance, the interactions
of the mirror matter with the ordinary matter disk, and
also to some extent on initial conditions. 
The total amount of mirror matter required does not seem unreasonable, 
however. It corresponds to 
about $few \ \times 10^{5}$ mirror Hydrogen atoms per cubic centimetre
(or equivalent). If the mirror gas/dust
is spherically distributed with a radius of order 100 AU, then the total
mirror mass would be about that of a small planet ($\approx
10^{-6} M_{\odot}$) with only about $10^{-8} M_{\odot}$
within the orbit of Uranus which is about two orders
of magnitude within present limits\cite{limits}.  If the configuration is 
disk-like rather than spherical, then the total mass of mirror
matter would obviously be even less. The requirement 
that the mirror gas/dust be
denser than its ordinary counterpart at these distances could be due
to the ordinary material having been expelled by solar pressure.

Fortunately the uncertainty over the mirror matter configuration
does not preclude an experimental test of
our hypothesis.
It could be tested if another spacecraft moving with a different 
speed or with a different 
ratio of cross-sectional area to mass were used on
a future mission. The Cassini mission already launched in 
1997 and due to reach Saturn in July 2004 and also the 
proposed Pluto/Jupiter mission may provide suitable tests. 
Also, our hypothesis would be
falsified if an anomalous acceleration were observed to be
directed radially inward for a spacecraft that was {\it not} travelling
radially outward.

\vskip 0.4cm
\noindent
{\bf Acknowledgement}
\vskip 0.4cm
\noindent
RF is an Australian Research Fellow. RRV is funded 
by the Australian Research Council
and the University of Melbourne. Part of this work was done 
while RRV was in the
hospitable surroundings of Universit\`{a} di Roma I ``La Sapienza''.
We would like to thank Michael Nieto for patiently answering
some of our questions.

\end{document}